\documentclass[prb,twocolumn,amsmath,amssymb,floats,showpacs,floatfix]{revtex4-2}
\usepackage{graphicx}
\usepackage{bm}
\usepackage{amsmath}
\usepackage{amsfonts}
\usepackage{amsbsy}
\usepackage{physics}
\usepackage{verbatim}
\usepackage{color}
\usepackage{soul}
\usepackage{float}
\usepackage[nameinlink]{cleveref}
\crefmultiformat{figure}{~#2#1#3}{,\!~#2#1#3}{, #2#1#3}{ and~#2#1#3}
\crefname{figure}{}{}
\usepackage{placeins}

\begin{document}

\title{Synchronizing microwave cQED limit-cycle oscillators}

\author{Cecilie Hermansen$^{1}$}
\author{Jens Paaske$^{1}$}

\affiliation{$^{1}$Center for Quantum Devices, Niels Bohr Institute, University of Copenhagen, 2100 Copenhagen, Denmark}

\date{\today}
\begin{abstract}
Self-sustained oscillators play a central role in the stabilization and synchronization of complex dynamical systems. A number of different physical systems are currently being investigated to clarify the importance of such active components in the quantum realm. Here we explore the properties of a driven dissipative electron-photon hybrid system based on superconducting microwave resonators coupled resonantly to a voltage-biased double quantum dot (DQD). First, we establish a Hopf bifurcation at a critical value of the electron-photon coupling, beyond which an effective negative friction sustains steady limit-cycle oscillations of individual resonators. Second, we show that two such limit-cycle resonators coupled via the same voltage-biased DQD synchronize for small enough frequency detuning. A nonlinear photon Keldysh action is derived by perturbation theory in the effective circuit fine-structure constant, and the limit-cycle dynamics is analyzed in terms of resulting saddle-point, and Fokker-Planck equations. In the Markovian limit of infinite bias voltage, these results are shown to agree well with the solution of a corresponding Lindblad master equation for the DQD resonator system.
\end{abstract}

\maketitle

\section{Introduction}

Limit-cycle oscillators play a central role in nature, underlying important phenomena such as stabilization and synchronization crucial to a wide range of dynamical systems~\cite{Strogatz2024, Pikovsky2001Oct, Minorsky1947, Balanov2009}. Recent theoretical works have explored the role of limit-cycles in open quantum mechanical systems~\cite{Marquardt2009May, Lorch2014Jan, Chia2020Oct, BenArosh2021Feb, Sudler2024May, Dutta2025Feb}, and a large body of work~\cite{Mari2013Sep, Lee2013Dec, Walter2014Mar, Walter2015Jan, Lorch2016Aug, Tindall2020Jan, Buca2022Mar, Kehrer2024Oct} has been devoted to the study of quantum synchronization in both continuous variable quantum oscillators and discrete quantum systems. 

The paradigmatic van der Pol oscillator~\cite{vdPol1926Nov, Ginoux2012Apr} has had an enormous impact on nonlinear dynamics in a wide range of fields ranging from cardiology and neurology to robotics and active matter. The exemplary physical system studied by van der Pol a century ago was a simple lumped element resonator with nonlinear feedback from a triode. Now, one century later, it is therefore natural to inquire about the miniaturization of such a non-linear electronic circuit into the quantum realm, and its potential for quantum limit-cycle dynamics. Triodes have since been replaced by operational amplifiers, and the obvious quantum replacement component would appear to be the single-electron transistor in the form of gateable semiconductor quantum dots.

In 2012, Frey et al.~\cite{Frey2012Jan} realized a hybrid quantum device, in which an electrically gateable semiconductor double quantum dot (DQD) was demonstrably dipole coupled to the microwave field of a superconducting coplanar waveguide resonator. As pointed out also by Petersson et al.~\cite{Petersson2010Aug}, the quantum capacitance, reflecting the gate dependence of the internal electronic kinetic energy of the DQD, allowed the resonator to pick up a signal from a capacitive coupling to one of the quantum dots, even when pinching off the external tunnel barriers. Later, a pair of voltage-biased DQDs were shown to lead to microwave masing above a certain threshold bias~\cite{Liu2014Jul, Liu2015Jan, Liu2015Nov, Liu2017Aug}. With the continuous development of high-impedance superconducting resonators, the DQD-resonator hybrid system has since been pushed to the strong coupling regime~\cite{Stockklauser2015Jul, Mi2016Dec, Stockklauser2017Mar, Bruhat2018Oct, Scarlino2019May, Scarlino2022Jul, Ungerer2024Feb} and an earlier proposal for distant microwave mediated coupling of two DQDs has been realized~\cite{Childress2004Apr, vanWoerkom2018Oct}. The experimental status in the field as of 2017 has been nicely summarized in Refs.~\onlinecite{Cottet2017Sep, Janik2025Mar}, where also more references are provided.

\begin{figure}[t]
\centering
\includegraphics[width=\linewidth]{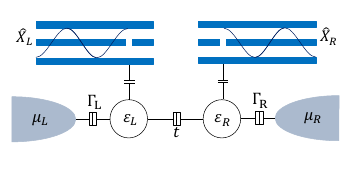}
\caption{Voltage-biased double quantum dot with each dot coupled capacitively 
to individual superconducting single-mode microwave resonators.}
\label{fig:DQD_sketch}
\end{figure}
Here, we revisit the basic DQD-resonator system (cf. Fig.~\ref{fig:DQD_sketch}) and carry out a theoretical analysis of its non-equilibrium dissipative nonlinear quantum dynamics. Starting from a microscopic model, we derive an effective Keldysh action for the resonator photons, including their (quartic) nonlinearities deriving from the voltage-biased DQD. From this we confirm the observed Hopf bifurcation to a stable quantum limit cycle beyond a critical electron-photon coupling, consistent with the experimentally demonstrated maser action~\cite{Liu2014Jul, Liu2015Jan, Liu2017Aug} and its demonstrated entrainment to an external drive~\cite{Liu2015Nov, Liu2017Nov}. We find that a single resonator coupled to a voltage-biased DQD realizes a {\it quantum Stuart-Landau oscillator} with a bifurcation to a simple elliptic limit cycle in phase space, and find no evidence for genuine quantum van der Pol relaxation oscillator with non-elliptical phase space orbits and two characteristic time scales~\cite{Chia2020Oct, BenArosh2021Feb, Sudler2024May}. Coupling two resonators with slightly different frequencies to the same DQD device (cf. Fig.~\ref{fig:DQD_sketch}), we find evidence for quantum synchronization. Already the linear theory shows a characteristic level-attraction but becomes unstable as the electron-photon coupling is increased through the bifurcation. The stable nonlinear theory shows clear evidence for phase and frequency synchronization of two limit-cycle oscillators.

The system is analyzed in terms of an effective Keldysh action for the photons, giving rise to stable saddle-point equations and a corresponding Fokker-Planck equation describing the fluctuations. We also include the results of a Lindblad Master Equation (LME), which is solved using the QuTip Python Package~\cite{johansson2012qutip, johansson2013qutip, lambert2026qutip5}. This is straightforward to do in the infinite-bias case, where the system is Markovian, whereas the case of a finite bias voltage is treated within the approximate PERLind approach~\cite{Kirsanskas2018Jan}. We have applied both methods to a similar (voltage-biased triple-dot) system in Ref.~\onlinecite{Hermansen2024Nov}, and refer to this paper for more details on their implementation.

\section{Model}
We consider the device sketched in Fig.~\ref{fig:DQD_sketch} comprising a voltage-biased DQD, with one microwave resonator capacitively coupled to each QD ($i=L/R$), described by Hamiltonian $H= H_\text{el}+ H_\text{ph}+H_\text{el-ph}$ with
\begin{align}\label{eq:DQD_Hamiltonian}
H_\text{el}=&\sum_{i}\varepsilon_i d_i^\dagger d_i + t(d_L^\dagger d_R +d_R^\dagger d_L)\\
&+\sum_{i,{\mathbf k}}\xi_{i{\mathbf k}}^{} c_{i{\mathbf k}}^\dagger c_{i{\mathbf k}}+\!\!\sum_{i,{\mathbf k}}\!\left(t_i c_{i{\mathbf k}}^\dagger d_i +\text{h.c.}\right)\nonumber\\
H_\text{ph}=&\,\sum_{i}\omega_i a_i^\dagger a_i  + \sum_{i,s} \omega_{is} a^\dagger_{is} a_{is}\\
&+\!\!\sum_{i,s} g_{is} (a_i^\dagger +a_i)(a_{is}^\dagger +a_{is})\nonumber\\
H_\text{el-ph}=&\sum_{i} g_i d_i^\dagger d_i (a_i^\dagger+a_i).   
\end{align}
Each QD ($d^\dagger_{i}$) is coupled to a normal lead ($c^\dagger_{i k}$), and each resonator ($a^\dagger_{i}$) is coupled to a photonic environment with a continuum of modes, $a^\dagger_s$.
The QD-resonator coupling constant is given as $g_{i}=\omega_{i}v_{g}\sqrt{\pi Z/R_{Q}}$, in terms of the resonator impedance $Z$, the resistance quantum $R_{Q}=2\pi\hbar/e^{2}=25.8\,{\text k}\Omega$, and a capacitive lever arm $v_{g}=\sqrt{C_{g}/(C_{g}+C_{\Sigma})}$, expressed in terms of a coupling capacitance and the total capacitance of the QD to other gates~\cite{Childress2004Apr}. We shall neglect Coulomb interactions on the DQD altogether and consider only spinless electrons. As demonstrated in Appendix~\ref{app:Coulomb}, this omission does not implicate any qualitative changes in the main conclusions drawn below, which rely mainly on the energy transfer to the resonator taking place via sequential tunneling at large bias voltages.  

Throughout the paper, we will restrict our attention to a specific exemplary parameter set, for which the resonator is resonant with the electronic excitation energy of the DQD, $\omega_0=2\sqrt{[(\varepsilon_{L}-\varepsilon_{R})/2]^{2}+t^2}$. In the case of a single mode, unless otherwise specified, we always use $\omega_0=1,\, g_L=0,\, g_R=g,\, \kappa=0.001\omega_0,\, \varepsilon/2=\varepsilon_L=-\varepsilon_R=(\sqrt{3}/4)\omega_0,\, t=\omega_0/4,\, \Gamma_L=\Gamma_R=0.05\omega_0,\, T=0.001\omega_{0}$ and $V=10 \omega_0.$ In the case of two modes we allow for a finite frequency detuning, $\omega_{L/R}=\omega_0\pm\delta\omega/2$, and assume equal coupling strengths, $g_L=g_R=g$. Except where explicitly stated otherwise, we work in units where $\hbar=k_{B}=1$.

\subsection{Keldysh action}

To assess the dynamics of this open quantum system driven out of equilibrium, we construct the partition function as the coherent-state path integral on the Keldysh contour
\begin{equation}
Z=\int\mathcal{D}[\bar{\phi},\phi,\bar{\psi},\psi]e^{i\left(S_{\text{el}}[\bar{\psi},\psi]+S_\text{ph}[\bar{\phi},\phi,]+S_\text{el-ph}[\bar{\phi},\phi,\bar{\psi},\psi]\right)},
\label{eq:partition_function}
\end{equation}
in terms of complex boson field $\phi_{i}^{\alpha}$, Keldysh rotated to classical, and quantum components $\phi_i^\text{cl/q}=\left(\phi_i^+\pm \phi_i^-\right)/\sqrt{2}$, and a Grassmann field $\psi_{ia}$ for the QD electrons, with Larkin-Ovchinnikov (LO) component indices $a=1,2$~\cite{Kamenev2023Jan}. The Keldysh action corresponds to each of the three terms in Eq.~\eqref{eq:DQD_Hamiltonian} after integrating out the lead electrons and the bosonic baths:
\begin{align}
S_\text{el}=&\int^\infty_{-\infty}\!\!\!\dd t\int^\infty_{-\infty}\!\!\!\dd t'\, \bar{\psi}_{i}(t)\hat{G}^{-1}_{ij}(t-t')\psi_{j}(t'),\\
S_\text{ph}
    =&\int^\infty_{-\infty}\!\!\!\dd t\int^\infty_{-\infty}\!\!\!\dd t'\,\bar{\phi}_{i}(t)\hat{D}_{0,ij}^{-1}(t-t')\phi_{j}(t'),\\
S_\text{el-ph}&=-\int_{-\infty}^\infty\!\!\dd t\,g_i \bar{\psi}_{ia}(t)\psi_{ib}(t)\hat{\gamma}_{ab}^{\alpha}\left[\bar{\phi}_{i}^{\alpha}(t)+\phi_{i}^{\alpha}(t)\right],
\end{align}
with Keldysh interaction vertex tensors given in terms of Pauli matrices as $\hat{\gamma}_{ab}^{\text{cl}}=\hat{\sigma}^{0}_{ab}/\sqrt{2}$ and $\hat{\gamma}_{ab}^{\text q}=\hat{\sigma}^{1}_{ab}/\sqrt{2}$. The electronic Keldysh matrix Green function reads
\begin{align}
\hat{G}_{ij}(t-t')=\begin{pmatrix} G^\text{R}_{ij}(t-t') && G^\text{K}_{ij}(t-t') \\0 &&G^\text{A}_{ij}(t-t')
\label{eq:general_fermion_keldysh_GF}
\end{pmatrix},
\end{align}
with inverse component ($L/R$-matrix) Green functions given in the wide-band limit, respectively, by the Fourier transforms
\begin{align}
\left(G^\text{R/A}(\omega)\right)^{-1}_{ij}=&\begin{pmatrix}
\omega-\varepsilon_L\pm i\Gamma_L && -t\\-t && \omega -\varepsilon_R\pm i \Gamma_R
\end{pmatrix}_{ij}\nonumber\\
\quad \left(G^{-1}(\omega)\right)^\text{K}_{ij}=&\begin{pmatrix}
2i\Gamma_L F_{L}(\omega) && 0\\0&& 2i\Gamma_R F_{R}(\omega)
\end{pmatrix}_{ij}.\end{align}
The tunneling rates $\Gamma_i=\pi\rho_i|t_i|^2$ are expressed in terms of the constant density of states in the leads and a momentum-independent tunnel coupling, and the distribution function $F_{L/R}(\omega)=\tanh((\omega-\mu_{\alpha})/2T)$ encodes the temperature of the two leads and their chemical potentials, $\mu_{L}-\mu_{R}=eV$, differing by the applied bias-voltage. Both resonators are coupled to independent bosonic baths, which are assumed to have an ohmic density of states,
$J_{i}(\omega) = 4\kappa_{i}\omega$, and are maintained at thermal equilibrium with the
same temperature, $T$, as the electron system. Under these assumptions, and assuming that $\kappa_{i}\ll\omega_{i}$, the corresponding photon Green functions read 
\begin{align}
\hat{D}_{0,ij}(t-t')=\begin{pmatrix} D_{0,ij}^\text{K}(t-t') && D_{0,ij}^\text{R}(t-t') \\D_{0,ij}^\text{A}(t-t') && 0
\label{eq:general_fermion_keldysh_GF}
\end{pmatrix},
\end{align}
with Fourier transformed components
\begin{align}
D^\text{R/A}_{0,ij}(\omega&)=\frac{\delta_{ij}}{\omega-\omega_i\pm i\kappa_{i}},\nonumber\\
D^\text{K}_{0,ij}(\omega)&=\delta_{ij}\coth\left(\frac{\omega_{i}}{2T}\right)\left[D_{0,ii}^\text{R}(\omega)-D_{0,ii}^\text{A}(\omega)\right].
\end{align}

Below, we shall refer also to the dimensionless photon displacement field, $X_i^\alpha(t)=[\bar{\phi}_i^\alpha(t)+\phi_i^\alpha(t)]/\sqrt{2}$, and momentum, $P_{i}^{\alpha}(t)=i[\bar{\phi}_i^\alpha(t)-\phi_i^\alpha(t)]/\sqrt{2}$, with $\alpha=$cl,q and $i=L,R$. Reinstating $\hbar$, these fields translate to actual resonator charge and flux as
$Q(t)=X(t)q_{0}$ and $\Phi(t)=P(t)\hbar/q_{0}$, in terms of the zero-point fluctuation charge, $q_{0}=\sqrt{\hbar/Z}=e\sqrt{R_{Q}/(2\pi Z)}$.

\subsection{Effective photon action to fourth order in $g$}

An effective nonlinear action for the resonators may be obtained by integrating out the QD electrons. In the case of interest, where the electronic system is resonant with the microwave cavity, there is no good separation of energy scales to allow for an adiabatic approximation~\cite{Mozyrsky2006Jan, Schiro2014May}. Instead, we shall therefore carry out a perturbative expansion in $g_{i}/\omega_{i}\ll 1$, including up to fourth order terms in an expansion of the resulting $\tr\log$ term, i.e. second order in the effective fine-structure constant, $\alpha_{\rm eff}\equiv(g_{i}/\omega_{i})^{2}=2\pi v_{g}^{2}(Z/Z_{0})\alpha$, where $\alpha\approx 1/137$ denotes the vacuum fine-structure constant and $Z_{0}\approx 377\,\Omega$ the vacuum impedance.

In order to proceed with this effective nonlinear photon action, we shall assume that the photon spectral functions are sharply peaked compared to the electron spectral functions, i.e. $\kappa\ll\Gamma,$ and restrict ourselves to considering resonator detunings smaller than the width of the electron spectral function, i.e. $|\delta\omega|\ll\Gamma$ with $\omega_{L/R}=\omega_{0}\pm\delta\omega/2$. In effect, this allows us to evaluate the electron Green functions with only on-shell photon frequencies, i.e. $G(\varepsilon+\omega)\approx G(\varepsilon+\omega_0)$ inside the frequency integrals. These assumptions lead to a time-local effective action with no anomalous terms~\cite{Cottet2020Oct}:
\begin{align}\label{eq:Seff}
S_\text{eff}[\bar{\phi},\phi]&\approx S_\text{ph}[\bar{\phi},\phi]+\sum_{i=1}^{4} S^{(i)}[\bar{\phi},\phi].
\end{align}
The first order term yields
\begin{align}
S^{(1)}=-\sqrt{2}\sum_i g_i n_i \int\!\dd t \left[\bar{\phi}_{i}^{\text q}(t)+\phi_{i}^{\text q}(t)\right],
\end{align} 
with QD occupation given by $n_i=-i\int\frac{\dd\omega}{2\pi}G^<_{ii}(\omega).$ This term can therefore be removed from the action by a shift of the classical field, $\phi^\text{cl}_i\rightarrow \phi^\text{cl}_i-g_i n_i/\omega_i,$ and reintroduced before calculating actual photon expectation values.

\begin{figure}[t]
    \centering
    \includegraphics[width=\linewidth]{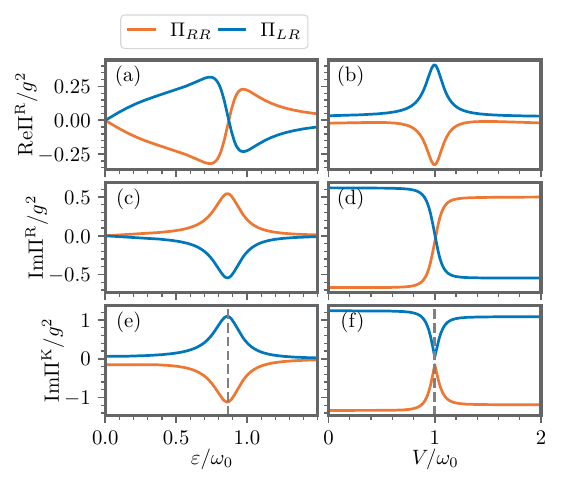}
    \caption{Real and imaginary parts of the retarded, and imaginary part of the Keldysh photon self energy plotted in panels (a, c, e) against DQD detuning for $V=10\omega_0$, and in panels (b, d, f) against bias voltage for $\varepsilon=(\sqrt{3}/2)\omega_{0}$. In both cases our $L/R$-symmetric parameter set implies that $\Pi^\text{R}_{LL}=\Pi^\text{R}_{RR}$ and $\Pi^\text{R}_{LR}=\Pi^\text{R}_{RL}.$ The dashed gridline in panel (e) shows the resonance condition, $\varepsilon=(\sqrt{3}/2)\omega_{0}$, and in (f) the threshold condition for the bias voltage, $V=\omega_0.$ The width of the transition region around the dashed gridline is set by $\Gamma$.}
    \label{fig:polarization}
\end{figure}

The second order term gives rise to a self-energy, which mixes the $i=L,R$ terms in the quadratic part of the photon action:
\begin{align}
S^{(2)}=-\int dt \bar{\phi}^\alpha_i(t) \hat{\Pi}^{\alpha\beta}_{ij}\phi^\beta_j(t),
\end{align}
where the electronic polarization bubble has been evaluated at frequency $\omega_{0}$ and is found as
\begin{align}
\Pi^\text{R/A}_{ij}&=-\frac{i g_i g_j}{2}
\int\frac{d\varepsilon}{2\pi}\left(G^\text{K}_{ij}(\varepsilon)G_{ji}^\text{A/R}(\varepsilon-\omega_0)\right.\\
&\hspace{6mm}\left.+G_{ij}^\text{R/A}(\varepsilon)G_{ji}^\text{K}(\varepsilon-\omega_0)\right)\nonumber\\ 
\Pi^\text{K}_{ij}&=-\frac{i g_i g_j}{2} \int \frac{d\varepsilon}{2\pi}\left(G_{ij}^\text{R}(\varepsilon)G^\text{A}_{ji}(\varepsilon-\omega_0)\right.\\
&\hspace{6mm}\left.+G_{ij}^\text{A}(\varepsilon)G_{ji}^\text{R}(\varepsilon-\omega_0)+G_{ij}^\text{K}(\varepsilon)G_{ji}^\text{K}(\varepsilon-\omega_0)\right).\nonumber
\end{align}
These time-independent complex photon self-energies are plotted in Fig.~\ref{fig:polarization} as a function of either DQD detuning or bias voltage.

The third order term, $S^{(3)}$, vanishes within this approximation since not all three photon fields can be on-shell at the same time. 

\begin{figure}[t]
    \centering
    \includegraphics[width=\linewidth]{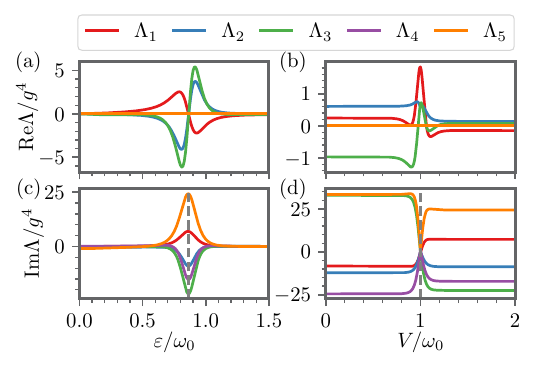}
    \caption{Real and imaginary parts of the photon interaction vertices appearing in the fourth order contribution to the effective action plotted in panels (a, c) against DQD detuning for $V=10\omega_0$, and in panels (b, d) against bias voltage for $\varepsilon=(\sqrt{3}/2)\omega_{0}$. The dashed gridline in (c) shows the resonance condition, $\varepsilon=(\sqrt{3}/2)\omega_{0}$, and in (d) the threshold condition for the bias voltage, $V=\omega_0.$ The width of the transition region around the dashed gridline is set by $\Gamma$.}
    \label{fig:lambda}
\end{figure}

The fourth order term yields
\begin{equation}
S^{(4)}=-\int \dd t\, \bar{\phi}_i^{\alpha}(t)\bar{\phi}_j^{\beta}(t)\hat{\Upsilon}^{\alpha\beta\gamma\delta}_{ijkl}\phi_k^{\gamma}(t)\phi_l^{\delta}(t),
\end{equation}
with the interaction tensor obtained as 
\begin{align}
\hat{\Upsilon}^{\alpha\beta\gamma\delta}_{ijkl}&=-\frac{ig_ig_jg_kg_l}{4}\Tr\int\frac{\dd\varepsilon}{2\pi}\\
&\hspace{-8mm}\Bigg[G_{li}(\varepsilon)\gamma^\alpha G_{ij}(\varepsilon+\omega_0)\gamma^\beta G_{jk}(\varepsilon+2\omega_0)\gamma^\gamma G_{kl}(\varepsilon+\omega_0)\gamma^\delta\nonumber\\
&\hspace{-3mm}+\frac{1}{2}G_{li}(\varepsilon)\gamma^\alpha G_{ij}(\varepsilon+\omega_0)\gamma^\gamma G_{jk}(\varepsilon)\gamma^\beta G_{kl}(\varepsilon+\omega_0)\gamma^\delta\Bigg]\nonumber.
\end{align}
For the single-mode case ($g_L=0, g_R=g$) this is conveniently written in terms of $\phi\equiv\phi_R$ as~\cite{Cottet2020Oct}, 
\begin{equation}
S^{(4)}=-\int\!\dd t\!\begin{pmatrix}
        \bar{\phi}^\text{cl}\bar{\phi}^\text{cl}\\  \bar{\phi}^\text{cl}\bar{\phi}^\text{q}\\ \bar{\phi}^\text{q}\bar{\phi}^\text{q}
    \end{pmatrix}^{T}\!\!
    \begin{pmatrix}
    0 && \Lambda_1 &&\Lambda_2 \\\Lambda_1^* && \Lambda_5 &&\Lambda_3\\-\Lambda_2^* && \Lambda_3^* &&  \Lambda_4  
    \end{pmatrix}
    \begin{pmatrix}
        \phi^\text{cl}\phi^\text{cl} \\ \phi^\text{cl}\phi^\text{q}\\ \phi^\text{q}\phi^\text{q}
    \end{pmatrix},
    \label{eq:fourth_order_action_single_mode}
\end{equation}
where
\begin{align}
\Lambda_1&=\hat{\Upsilon}^\text{ccqc}+\hat{\Upsilon}^\text{cccq}\nonumber\\
\Lambda_2&=\hat{\Upsilon}^\text{ccqq}\nonumber\\
\Lambda_3&=\hat{\Upsilon}^\text{qcqq}+\hat{\Upsilon}^\text{cqqq}\label{eq:naming_fourth_order}\\
\Lambda_4&=\hat{\Upsilon}^\text{qqqq}\nonumber\\
\Lambda_5&=\hat{\Upsilon}^\text{cqcq}+\hat{\Upsilon}^\text{cqqc}+\hat{\Upsilon}^\text{qcqc}+\hat{\Upsilon}^\text{qccq}\nonumber.
\end{align}
All photon interaction vertices, except $\Lambda_{5}$, are labeled by the number of quantum fields involved in the interaction. These time-independent complex interaction vertices are plotted in Figs.~\ref{fig:lambda} as functions of either DQD detuning or bias voltage.

\section{Single mode results}

\subsection{Saddle point equations}

In the single-mode case, four saddle point equations are obtained as
\begin{equation}
\frac{\delta S}{\delta \bar{\phi}^\text{q}_i}=
\frac{\delta S}{\delta \phi^\text{q}_i}=
\frac{\delta S}{\delta \bar{\phi}^\text{cl}_i}=
\frac{\delta S}{\delta \phi^\text{cl}_i}=0.
\end{equation}
As noted also in Refs.~\onlinecite{Kamenev2023Jan, Thompson2022Apr} for other systems, these four equations do not come in pairs of complex conjugate equations. Nevertheless, this is readily rectified by transforming the quantum field as $\chi=-\bar{\phi}^\text{q}$ and $\bar{\chi}=\phi^\text{q}$, whereby, noting that $\Lambda_{4}$ and $\Lambda_{5}$ are both purely imaginary, one arrives at the following two coupled saddle-point equations and their complex conjugates:
\begin{align}
\!\!\partial_t \phi &=-(i\tilde{\omega}_{0}+\tilde{\kappa}+i\Lambda_{1}^{\ast}|\phi|^2+2i\Lambda_{3}^{\ast}|\chi|^{2})\phi-2i\Lambda_{2}^{\ast}\phi^{2}\chi\nonumber\\
    &\hspace*{4mm}-(D_{0}+i\Lambda_{5}|\phi|^2)\bar{\chi}+2i\Lambda_{4}|\chi|^{2}\bar{\chi}-i\Lambda_{3}\bar{\chi}^{2}\bar{\phi},\nonumber\\ 
\!\!\partial_t \chi &=(i\tilde{\omega}_{0}+\tilde{\kappa}-i\Lambda_{3}^{\ast}|\chi|^{2}+2i\Lambda_{1}^{\ast}|\phi|^2)\chi+2i\Lambda_{2}^{\ast}\chi^{2}\phi\nonumber\\
    &\hspace*{4mm}+i\Lambda_{5}|\chi|^{2}\bar{\phi}-i\Lambda_{1}\bar{\phi}^{2}\bar{\chi},
   \label{eq:saddle_point_single_mode}
\end{align}
where we have dropped the $\text{cl}$ index from the classical field, $\phi$, dressed the frequency and the damping rate as $\tilde{\omega}_{0}=\omega_{0}+\Pi'$ and  $\tilde{\kappa}=\kappa-\Pi''$, where $\Pi_{RR}^\text{R}=\Pi'+i\Pi''$, and we have collected the Keldysh-component self-energies into $D_{0}=2\kappa\coth(\omega_{0}/2T)+i\Pi^\text{K}$. Setting $\chi=0$ this becomes the equations of motion for the Stuart-Landau oscillator~\cite{Stuart1960Nov, Landau1944, Pikovsky2001Oct}. Since Eqs.~\eqref{eq:saddle_point_single_mode} now come together with their complex conjugates, we may switch to a polar representation of the fields, $\phi=r e^{i\theta}$, whereby
\begin{align}
\dot{r}&=-\tilde{\kappa}r-\Lambda_{1}''r^3\\
\dot{\theta}&=-\tilde{\omega}_{0}-\Lambda_{1}'r^2.\label{eq:limitcyclefreq}
\end{align}
The resonator displays a trivial fixed point with $r=0$ as well as a limit cycle with $r=r_*=\sqrt{-\tilde{\kappa}/\Lambda_{1}''}$ and a running phase, $\theta(t)=-(\tilde{\omega}_{0}+\Lambda_{1}'r_*^2)t$ with the characteristic amplitude-dependent frequency softening of the Duffing oscillator~\cite{Strogatz2024}. Since $\Lambda_{1}''>0$ for $V>\omega_0$ (cf. Fig.~\ref{fig:lambda}), the Hopf bifurcation from fixed point to limit cycle takes place when $\tilde{\kappa}$ turns negative. For infinite bias voltage this happens when $\kappa<g^{2}\varepsilon t^{2}\sqrt{4t^{2}+\varepsilon^{2}}(4t^{2}+10\Gamma^{2}+\varepsilon^{2})(4t^{2}+4\Gamma^{2}+\varepsilon^{2})^{-2}(4t^{2}+\Gamma^{2}+\varepsilon^{2})^{-1}(2\Gamma)^{-1}$, i.e. $2\kappa\Gamma\lesssim g^{2}t^{2}\varepsilon/(4t^{2}+\varepsilon^{2})^{3/2}$ for $\Gamma\ll t$, as for the parameters explored here. In more physical terms, this condition reads $(4\kappa/\omega_{0})(4\Gamma/\varepsilon)\lesssim Z \omega_{0} C_{Q}(\varepsilon)$, in terms of the quantum capacitance for the DQD, $C_{Q}(\varepsilon)=e^{2}v_{g}^{2}4t(\varepsilon^{2}+4t^{2})^{-3/2}$~\cite{Petersson2010Aug, Ota2010Jan}. For the parameters used in Figs.~\ref{fig:polarization} and~\ref{fig:lambda}, this implies a critical coupling of $g_{c}/\omega_{0}=0.043$. For couplings larger than this value, a dc bias voltage slightly ($\sim\Gamma$) larger than the resonator frequency provides the steady energy supply to make the dressed non-linear resonator self-sustained. For these parameters, the limit cycle radius becomes $r_{\ast}=2.87$ in units of the oscillator length, corresponding to some $(r_{\ast}^{2}-1)/2\approx 3.6$ photons, and the resonator frequency is reduced by merely $6.7\times 10^{-5}\omega_{0}$ from $\Pi'$ and by $8.2\times 10^{-6}\omega_{0}$ from $\Lambda_{1}'$ and $r_{\ast}$.

\subsection{Fluctuations: Fokker-Planck and beyond}
\begin{figure}
    \centering
    \includegraphics[width=\linewidth]{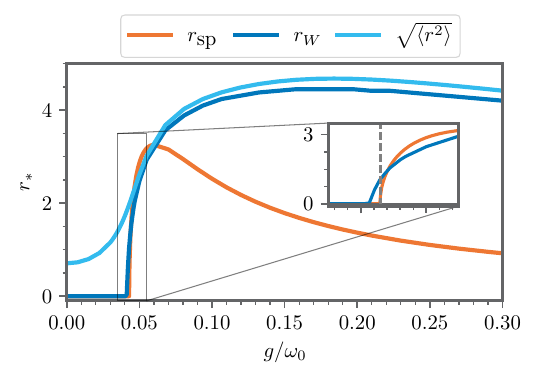}
    \caption{Limit cycle radius, $r_\ast$, determined respectively by the saddle-point equations, $r_\text{sp}$, or the value at which the Wigner function attains its maximum, $r_\text{W}$, and plotted together with the root mean square of $r$, calculated from the Wigner function. The inset shows a zoom in on the bifurcation region, $g\sim g_c$, revealing a slight difference in the predicted values for the critical coupling, $g_c$ (gridline marks $\tilde{\kappa}=0$).}
    \label{fig:limit_cycle_radius}
\end{figure}
Instead of solving Eqs.~\eqref{eq:saddle_point_single_mode}, one may follow the procedure outlined in Ref.~\onlinecite{Kamenev2023Jan} to derive a corresponding Fokker-Planck equation from the Keldysh action. Neglecting all nonlinearities except $\Lambda_{1}$, this leads to the equation
\begin{align}
\partial_t P &=\left[\,\,\,\partial_\phi \left(\tilde{\kappa}+i\tilde{\omega}_{0}+i\Lambda_1^{\ast}|\phi|^2\right)\phi\right.\label{eq:FP_equation_one_mode}\\
&\hspace*{5mm}\left.+\partial_{\bar{\phi}}\left(\tilde{\kappa}-i\tilde{\omega}_{0}-i\Lambda_1|\phi|^2\right)\bar{\phi}+D_0\partial_\phi\partial_{\bar{\phi}}\right]\!P\nonumber
\end{align}
for a distribution function, $P(\bar{\phi},\phi,t)$ and with diffusion constant $D_{0}$. Transforming to polar representation, $\phi=r e^{i\theta}$, and employing Wirtinger derivatives, $\partial_\phi=e^{-i\theta}(\partial_r-ir^{-1}\partial_\theta)/2$ and $\partial_{\bar{\phi}}=e^{i\theta}(\partial_r+ir^{-1}\partial_\theta)/2$, this transforms to
\begin{align}
\partial_t P&=\frac{1}{r}\partial_r(\tilde{\kappa}+\Lambda_1''r^2)r^2 P\nonumber\\
&\hspace{4mm}
+\frac{D_0}{4}\left(\partial_r^2+\frac{1}{r}\partial_r+\frac{1}{r^2}\partial_\theta^2\right)P\nonumber\\
&\hspace{4mm}+\left(\omega_0+\Pi'+\Lambda_1'r^2\right)\partial_\theta P,
\label{eq:radialeq1}
\end{align}
which allows for the $\theta$-independent steady-state solution,
\begin{equation}
    P_{0}(r)=ce^{-(2\tilde{\kappa}r^2+\Lambda_1''r^4)/D_0},
    \label{eq:single_mode_FP_solution}
\end{equation}
where $c$ ensures normalization, $\int_{0}^{\infty} \!dr\,2\pi r P_{0}(r)=1$, and the subscript $0$ indicates that $\Lambda_{2,3,4,5}$ are all omitted. The same solution was found in Ref.~\cite{Lee2013Dec}, for a LME with jump operators representing 1-photon drive and 2-photon losses. Without the nonlinearity, one can relate the distribution function to the Wigner function as $P(\bar{\phi},\phi,t)=W(\bar{\phi}/\sqrt{2},\phi/\sqrt{2},t)/2$. In the general nonlinear case, however, this identification is not strictly valid. In particular, we note that the solution~\eqref{eq:single_mode_FP_solution} is not a true Wigner function in the sense that the corresponding density matrix is not strictly positive semi-definite.

The solution~\eqref{eq:single_mode_FP_solution} is valid only if $\Lambda_1''/D_0>0$, which is the case for $V>\omega_{0}$ (cf. Fig.~\ref{fig:lambda}), in which case the Hopf bifurcation takes place as $\tilde{\kappa}$ passes through zero. Below and above the bifurcation, the maximum of the probability distribution is attained, respectively, at $r=0$, or $r=r_\ast=\sqrt{-\tilde{\kappa}/\Lambda_1''}$, consistent with the saddle point solution found above. This bifurcation is shown in Fig.~\ref{fig:limit_cycle_radius} as a sharp transition from the trivial fixed point, $r=0$, to the limit cycle at $r=r_\ast$ at a critical coupling, $g_{c}$, close to and above which $r_\ast\sim|g-g_{c}|^{1/2}$. The orange curve is determined as the fixed point of the classical saddle point equations, corresponding to the $r$ at which $P_{0}(r)$ reaches its maximum. The dark blue curve shows the exact radius of the limit cycle obtained as the radius at which the Wigner function attains its maximum. The Wigner functions is calculated using the LME, which treats the interaction between photons and electrons exactly. From the inset, the true bifurcation is seen to take place for a slightly smaller coupling than the value $g_c=0.043\omega_0$, determined from the saddle-point equations by the vanishing of $\tilde{\kappa}=0$.
Fig.~\ref{fig:limit_cycle_radius} also shows that the saddle-point value for the limit-cycle radius, $r_\text{sp}$, quickly becomes inaccurate for $g/\omega_0\gtrsim0.06.$

Including the effects of $\Lambda_{2}$ and $\Lambda_{5}$ adds multiplicative noise to the Fokker-Planck equation. With $\mathcal{L}_{1}P$ denoting the right hand side of Eq.~\eqref{eq:FP_equation_one_mode}, the resulting Fokker-Planck equation now takes the form, $\partial_{t}P=(\mathcal{L}_{1}+\mathcal{L}_{2})P$ with the additional contributions given by
\begin{align}
\mathcal{L}_{2}P=\frac{i\Lambda_2^*}{2}\partial_\phi^2\left(\phi^2P\right)-\frac{i\Lambda_2}{2}\partial_{\bar{\phi}}^2\left(\bar{\phi}^2P\right)+i\Lambda_5\partial_\phi\partial_{\bar{\phi}}\left(|\phi|^2P\right).
\label{eq:FP_equation_one_mode_nonlinear}
\end{align}
Assuming again that $P$ is independent of the angle $\theta$ of the polar representation, this corresponds to the following extra terms in the radial equation~\eqref{eq:radialeq1}:
\begin{align}
\mathcal{L}_{2}P=&\left(2\Lambda_2''+i\Lambda_5\right)P+
    \frac{1}{4}\left(7\Lambda_2''+5i\Lambda_5\right)r\partial_r P \nonumber\\
    &+\frac{1}{4}\left(\Lambda_2''+i\Lambda_5\right)r^2\partial_r^{2}P.
    \label{eq:radial_FP_with_nonlinear noise}
\end{align}
 We solve this eigenvalue problem numerically~\footnote{The calculations were performed using the Mathematica function NDeigensystem, where the area of the cells in the discrete spatial grid can be specified.}, and find that except for the zero eigenvalue corresponding to the steady state, all eigenvalues of the resulting differential operator $\mathcal{L}_{1}+\mathcal{L}_{2}$ are real and negative. 
 
\begin{figure}
    \centering
    \includegraphics[width=0.85\linewidth]{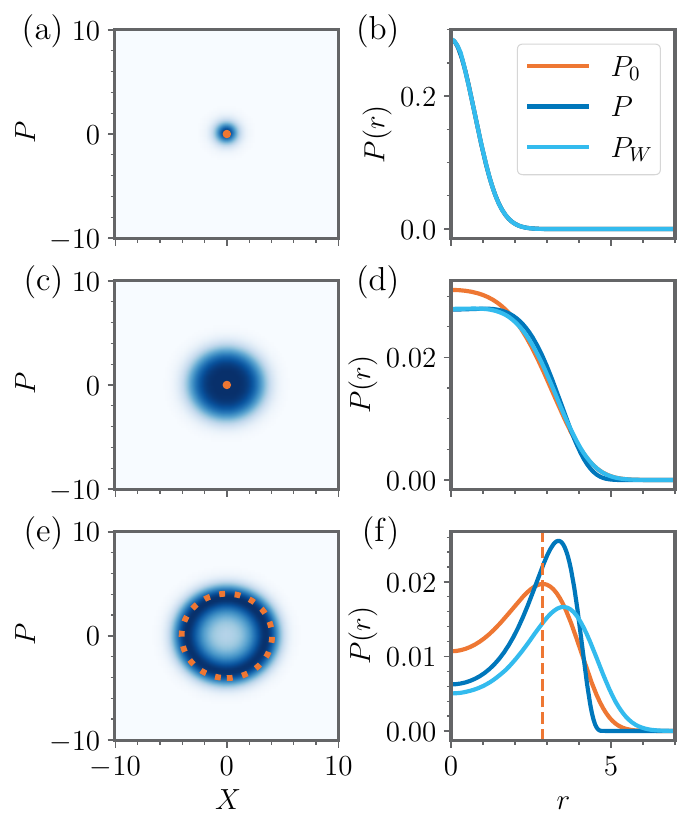}
    \caption{Wigner function (blue) overlaid by the limit-cycle radius, $r_{\ast}$, determined from the saddle-point equations (orange) [(a), (c), (e)] and radial probability distribution [(b), (d), (f)] obtained by solving the FP equation without/with ($P_0/P$) the nonlinear terms ($\Lambda_{2/5}$), plotted alongside $P_W=W(\bar{\phi}/\sqrt{2},\phi/\sqrt{2},t)/2$ with $\phi=r e^{i\theta}$. From top to bottom the coupling strength increases as $g/\omega_0=[0.01, 0.042, 0.05]$. The dashed gridline in panel (f) marks the maximum of $P_0$, i.e. the saddle-point value of the limit-cycle radius, $r_{\ast}$.}
    \label{fig:probability_distribution_comparison}
\end{figure}

 Fig.~\ref{fig:probability_distribution_comparison} shows a comparison of the steady state solution, $P$, to the previous solution, $P_{0}$, obtained from Eq.~\eqref{eq:single_mode_FP_solution}, and to the exact Wigner function, $W$, obtained from the LME. While all solutions preserve the bifurcation, Fig.~\ref{fig:probability_distribution_comparison}(d) shows that the inclusion of $\Lambda_{2}$ and $\Lambda_{5}$ lowers $g_{c}$ towards the correct value obtained from the Wigner function. Calculating the photon numbers as
 \begin{align}
 N_{\rm ph}=\pi\int\!\dd r\,r^{3}P_{i}(r)-1/2,
 \end{align}
the three different distributions, $P_{i}=\{P_0, P, P_{W}\}$, with $P_{W}=W(\bar{\phi}/\sqrt{2},\phi/\sqrt{2},t)/2$, shown in Figs.~\ref{fig:probability_distribution_comparison}(d) and (f), correspond to photon numbers $N_{\rm ph}=\{2.99, 2.85, 3.07\}$ and $N_{\rm ph}=\{4.54, 4.48, 6.34\}$, respectively. Despite their different shapes, both $P_0$ and $P$ agree very well on this lowest moment, which however disagrees markedly with that of the Wigner function already for $g=0.05$. Note that for the parameters considered here, the linear shift performed earlier, $\delta X^{\rm cl}=2gn_{R}/\omega_0$, is of the order of $10^{-2}$ and therefore is insignificant in these plots.

The two last terms, $\Lambda_{3,4}$, corresponding to photon interaction terms involving three, or four quantum fields give rise to third, and fourth order differential operators, acting as
\begin{align}
\mathcal{L}_{3}P=&\frac{i\Lambda_3}{2}\partial^2_{\bar{\phi}}\partial_\phi\left(\bar{\phi} P\right)-\frac{i\Lambda_3^{\ast}}{2}\partial_{\bar{\phi}}\partial^2_\phi\left(\phi P\right),\\
\mathcal{L}_{4}P=&-\frac{i\Lambda_4}{4}\partial^2_{\bar{\phi}}\partial^2_\phi P.
\end{align}
As seen from Fig.~\ref{fig:lambda}, these terms are comparable in magnitude to $\Lambda_{1,2,5}$, and there is a priori no reason to neglect them. In contrast, they are clearly part of the discrepancy between the Wigner function obtained from the LME and the Fokker-Planck distribution obtained without $\Lambda_{3,4}$, as well as terms of even higher order in $g$, which we neglect altogether. 
The only controlled way of truncating a Fokker-Planck equation is through a system-size expansion as described in Ref.~\onlinecite{Carmichael}, which results in a linear Fokker-Planck equation with only additive noise. However, since the qualitative agreement with the exact LME solution is already satisfactory, we shall pursue neither this refinement nor the quantitative influence of $\Lambda_{3,4}$ further in this work.

\subsection{Entrainment}

\begin{figure}
\centering
\includegraphics[width=\linewidth]{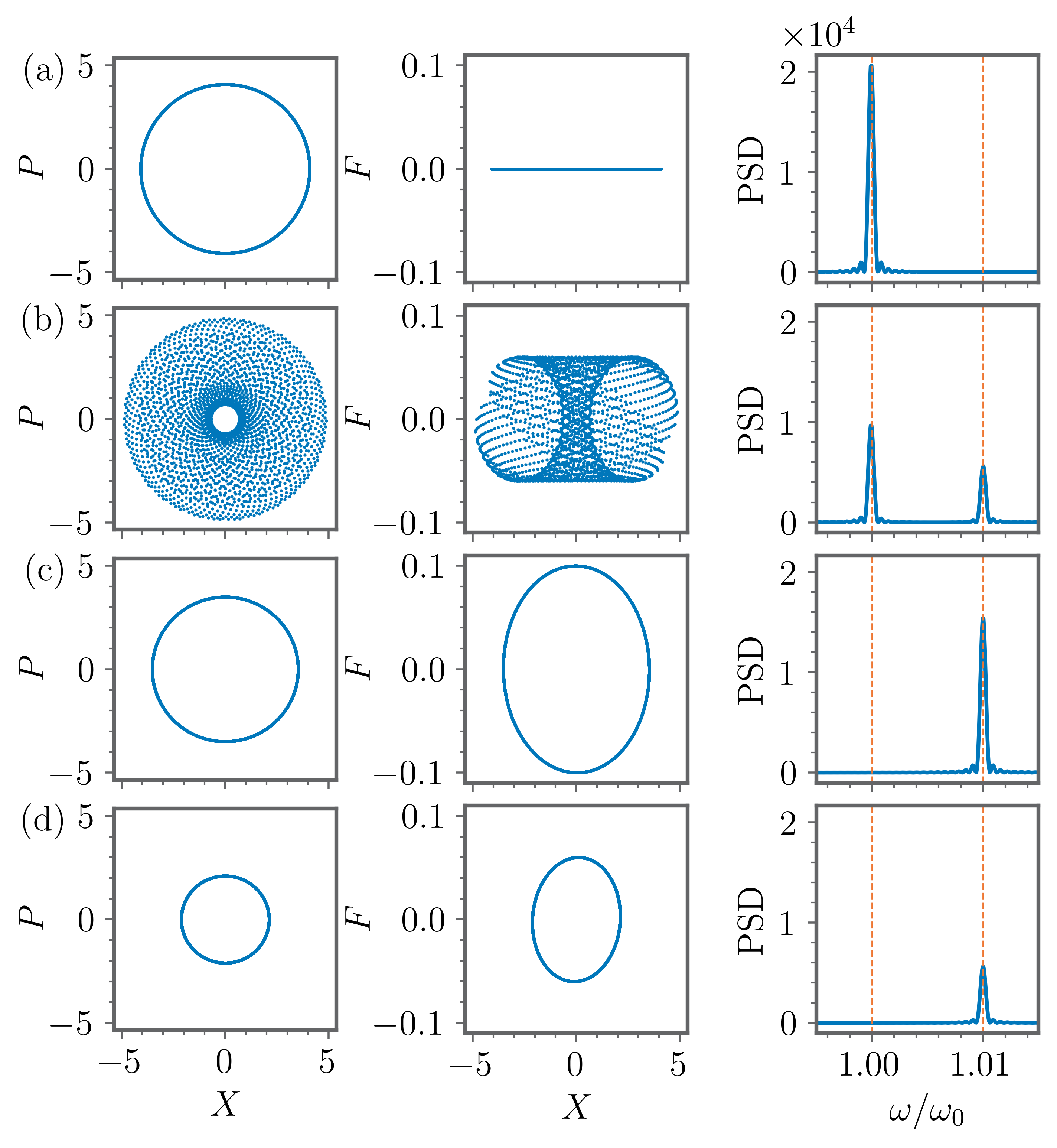}
\caption{Phase-space trajectory and Lissajous figure $(X(t), F(t))$ for a single resonator coupled to the DQD with external drive $F(t)=A\cos(\Omega t)$ and $\Omega=1.01\omega_0$. From (a) to (d) the amplitudes and the couplings are $A=[0, 0.06, 0.1,0.03]$ and $g/\omega_0=[0.05, 0.05, 0.05, 0.03]$.}
\label{fig:single_oscillator_Lissajous}
\end{figure}
As a final check on this effective non-linear oscillator, we demonstrate in Fig.~\ref{fig:single_oscillator_Lissajous}, that it can be entrained by an external harmonic force, $F(t)=A\cos(\Omega t)$, as long as its strength is large enough compared to its detuning, $|\Omega-\tilde{\omega}_{0}|$. Taking $\chi=0$ in Eqs.~\eqref{eq:saddle_point_single_mode} and adding an external force as:
\begin{align}
\!\!\partial_t \phi &=-(i\tilde{\omega}_{0}+\tilde{\kappa}+i\Lambda_{1}^{\ast}|\phi|^2)\phi+A\cos(\Omega t),
\end{align}
we solve for $(X(t),P(t))=({\rm{Re}}[\phi],{\rm{Im}}[\phi])\sqrt{2}$. The corresponding phase-space trajectories are shown in Fig.\ref{fig:single_oscillator_Lissajous} together with the Lissajous curves traced out by $(X(t),F(t))$, revealing frequency entrainment when the amplitude of the force exceeds a critical strength, found here to be $A_{c}=0.044$. The rightmost column shows the power spectral density (PSD), calculated from the auto-correlation function of $X(t)$ (cf. Appendix~\ref{app:psd}), which displays a clear transfer of spectral weight from $\tilde{\omega}_{0}$ to $\Omega$ as the amplitude of the force is increased. Note that the entrainment can also be found below the bifurcation point, $g<g_c.$ In this case any finite driving amplitude makes it possible to sustain a closed phase-space trajectory with finite radius, despite the fixed point of the undriven system being $r=0$ (see Fig.~\ref{fig:single_oscillator_Lissajous}(d)). The radius of the trajectory increases with increasing $A,$ and decreasing detuning, $|\Omega-\tilde{\omega}_0|$. This frequency entrainment is yet another signature of the dressed microwave resonator behaving as a self-sustained oscillator~\cite{Pikovsky2001Oct, Minorsky1947}.

\section{Two coupled modes}
Having established that a single microwave mode coupled to the voltage-biased DQD works as a self-sustained oscillator, we now consider the case of two modes as sketched in Fig.~\ref{fig:DQD_sketch}. In this case, the DQD provides for a driven-dissipative coupling between two self-sustained oscillators, which may be expected to synchronize~\cite{Kuznetsov2009Jul}. Already the linear model without the induced photon nonlinearities displays precursors to synchronization, and therefore we consider this case first.

\subsection{Eigenmodes of linear system}

\begin{figure}[t]
\centering
\includegraphics[width=\linewidth]{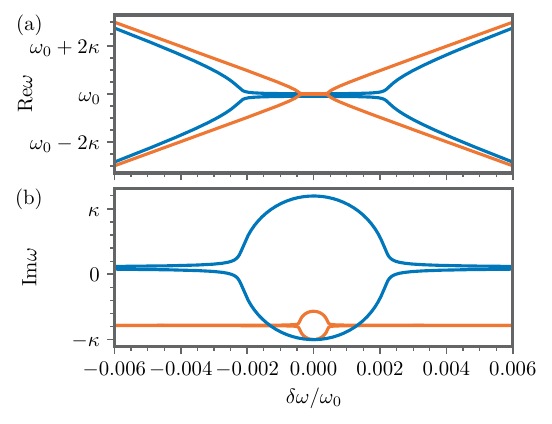}
\caption{Real (a) and imaginary (b) part of eigenfrequencies of the two-mode DQD-resonator system calculated in second-order perturbation theory. Orange(blue) lines are for $g=0.02$($g=0.045$) below(above) bifurcation.}
\label{fig:eigenmodes}
\end{figure}

The renormalized eigenfrequencies are readily found by solving the equation $\det\left((D_0^R)^{-1}(\omega)-\Pi^\text{R}\right)=0.$ The result is shown in Fig.~\ref{fig:eigenmodes} for two different values of the coupling strength. As a precursor to  synchronization, a pronounced level attraction~\cite{Bernier2018Aug, Lu2023Dec, Bourcin2024Dec} is observed in Fig.~\ref{fig:eigenmodes}(a). The approximate coincidence of eigenfrequencies persists through a region around zero detuning which grows with increasing $g$. Due to a small but finite value of the real parts of $\Pi^\text{R},$ the coincidence is not perfect and the system does not exhibit proper exceptional points demarcating the coincidence region. In line with the observations for a single mode, Fig.~\ref{fig:eigenmodes}(b) shows that one of the eigenfrequencies acquires a positive imaginary part when increasing the coupling beyond $g/\omega_0\approx 0.031$, corresponding to an apparent instability which, upon including nonlinearities, turns out to be a proper Hopf bifurcation.

The retarded photon self-energy, $\Pi^{R}_{ij}$, governing the linear analysis also informs about phase synchronization via the linear mode coupling it provides. As observed in Fig.~\ref{fig:polarization}, its real part is much smaller than its imaginary part close to resonance. For our $L/R-$symmetric parameter set, the diagonal and off-diagonal matrix elements of $\Pi^\text{R}$ differ only by a sign, $\Pi^\text{R}_{LL}=\Pi^\text{R}_{RR}=-\Pi^\text{R}_{LR}=-\Pi^\text{R}_{RL}\equiv i\Pi''$, and one arrives at the following approximate linear saddle-point equations corresponding to two  dissipatively coupled modes
\begin{align}
\partial_{t}\phi_L&=[-i(\omega_0+\delta\omega/2)-\kappa+\Pi'']\phi_L-\Pi''\phi_R,\notag\\
\partial_{t}\phi_R&=[-i(\omega_0-\delta\omega/2)-\kappa+\Pi'']\phi_R-\Pi''\phi_L.
\end{align}
Using polar coordinates, $\phi_{L/R}=r_{L/R}e^{i\theta_{L/R}}$, and assuming (as supported by the full non-linear analysis) both radii to be fixed and non-zero, this leads to the following Adler 
equation~\cite{Pikovsky2001Oct, Walter2015Jan} for the phase difference, $\theta=\theta_L-\theta_R$:
\begin{align}
\partial_{t}\theta=-\delta\omega + \Pi'' \left(\frac{r_L}{r_R}+\frac{r_R}{r_L}\right)\sin(\theta).
\label{eq:linear_phase_sync}
\end{align}
For zero detuning, this implies stable phase locking with $\theta=0$ for $\Pi''<0$, and $\theta=\pi$ for $\Pi''>0$, which is seen from Fig.~\ref{fig:polarization} to correspond respectively to bias voltage below, or above the resonator frequency. However, as finite limit-cycle radii are obtained only in the latter case ($V>\omega_{0}$), this implies $\theta=\pi$. 
For finite detuning, $\delta\omega\neq 0,$ the fixed-point value starts deviating from $\pi$ until $|\delta\omega|=|\Pi''(r_L/r_R+r_R/r_L)|$ (demarcating an Arnold tongue) where Eq.~\eqref{eq:linear_phase_sync} no longer has any stable solutions and $\theta$ becomes time dependent. This scenario is similar to the classical analysis provided in Ref.~\cite{Walter2015Jan}, when identifying $\Pi''$ with the dissipative coupling strength.

\subsection{Non-linear mode  coupling and synchronization}
\begin{figure*}[t]
    \includegraphics[width=0.8\linewidth]{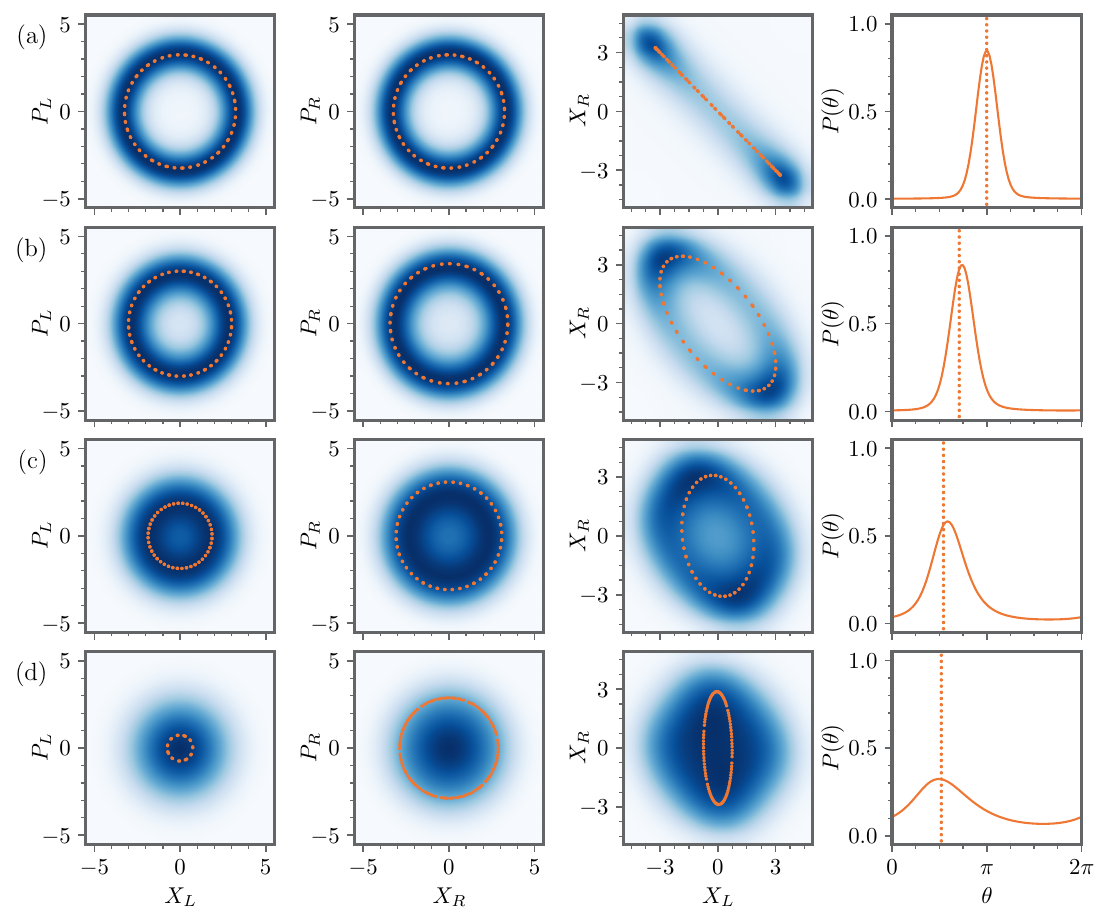}
    \caption{From left to right column: Wigner function for the left and right mode, joint position and phase difference probability distribution for $g_L=g_R=0.045$. The orange dotted lines are obtained from solutions of the steady-state saddle-point euqations. The blue density plots in the three leftmost, and the full orange line in the rightmost column correspond to solutions of the LME. From top to bottom row: (a) $\delta\omega=0$, (b) $\delta\omega=0.001$, (c) $\delta\omega=0.002$, and (d) $\delta\omega=0.004$.}
    \label{fig:two_mode_prob_dist}
\end{figure*}
Including the quartic terms from $\hat{\Upsilon}^{\alpha\beta\gamma\delta}_{ijkl}(\omega_{0})$ in the effective action, the corresponding classical saddle-point equations for the two modes are obtained from
\begin{equation}
0=\left.\frac{\delta S}{\delta \bar{\phi}^\text{q}_i}\right\vert_{\phi^\text{q}, \bar{\phi}^\text{q}=0}\!\!\!=\left.\frac{\delta S}{\delta \phi^\text{q}_i}\right\vert_{\phi^\text{q}, \bar{\phi}^\text{q}=0},
\end{equation}
which now contains the cubic nonlinearity
\begin{equation}
    \left.\frac{\delta S^{(4)}}{\delta \bar{\phi}^\text{q}_i}\right\vert_{\phi^\text{q}, \bar{\phi}^\text{q}=0}=-\sum_{jkl}\bar{\phi}^\text{cl}_j\phi^\text{cl}_k\phi^\text{cl}_l\left(\hat{\Upsilon}_{ijkl}^\text{qccc}+\hat{\Upsilon}_{jikl}^\text{cqcc}\right).
\end{equation}
The resulting equation of motion for $\phi_L^\text{cl}\equiv\phi_L$ reads:
\begin{align}
\partial_t &\phi_L=\notag\\
&-\left(i\omega_L+\kappa+i\Pi^R_{LL}\right)\phi_L-i\Pi^R_{LR}\phi_R\notag\\
&-i\left(\hat{\Upsilon}_{LLLL}^\text{qccc}+\hat{\Upsilon}_{LLLL}^\text{cqcc}\right)|\phi_L|^2\phi_L\nonumber\\
&-i\left(\hat{\Upsilon}_{LRRR}^\text{qccc}+\hat{\Upsilon}_{RLRR}^\text{cqcc}\right)|\phi_R|^2\phi_R
    \notag\\
&-i\left(\hat{\Upsilon}_{LLRL}^\text{qccc}+\hat{\Upsilon}_{LLRL}^\text{cqcc}+\hat{\Upsilon}_{LLLR}^\text{qccc}+\hat{\Upsilon}_{LLLR}^\text{cqcc}\right)|\phi_L|^2\phi_R\nonumber\\
&-i\left(\hat{\Upsilon}_{LRRL}^\text{qccc}+\hat{\Upsilon}_{RLRL}^\text{cqcc}+\hat{\Upsilon}_{LRLR}^\text{qccc}+\hat{\Upsilon}_{RLLR}^\text{cqcc}\right)|\phi_R|^2\phi_L\nonumber\\
&-i\left(\hat{\Upsilon}_{LRLL}^\text{qccc}+\hat{\Upsilon}_{RLLL}^\text{cqcc}\right)\bar{\phi}_R\phi_L^2\notag\\
&-i\left(\hat{\Upsilon}_{LLRR}^\text{qccc}+\hat{\Upsilon}_{LLRR}^\text{cqcc}\right)\bar{\phi}_L\phi_R^2\label{eq:saddle_point_full}.
\end{align}
The corresponding equation for $\bar{\phi}_L$ is obtained by taking the complex conjugate of Eq.~\eqref{eq:saddle_point_full}, and a corresponding set of equations for the right mode is obtained by interchanging $L\leftrightarrow R$. 
A numerical solution of these non-linear saddle-point equations yields the semiclassical phase-space trajectories in the absence of noise. These are shown as dashed lines in Fig.~\ref{fig:two_mode_prob_dist} for four increasing frequency detunings, $\delta\omega$, displaying two limit cycles with different radii. The corresponding dashed two-mode Lissajous figures, $(X_L,X_R)$, all reveal synchronization with a relative phase, which depends on detuning, $\delta\omega$. In the same panels of Fig.~\ref{fig:two_mode_prob_dist}, we show also the corresponding steady-state Wigner functions, calculated from the LME, which are largely consistent with the semiclassical phase portraits but including now the noise from quantum fluctuations. Obtaining the steady-state density matrix from solving the LME, the Lissajous figures are determined from the joint position probability distribution, calculated following Ref.~\cite{Walter2015Jan} as 
\begin{equation}
P(X_L, X_R)=\bra{X_L, X_R} \rho_{st} \ket{X_L, X_R},
\end{equation}
where $\ket{X_L, X_R}=\ket{X_L}\otimes \ket{X_R} $ is the tensor product of the eigenvectors of the position operator, $\hat{X}_{i}=(\hat{a}_{i}^\dagger+\hat{a}_{i})/\sqrt{2}$ with eigenvalue $X_{i}$ where $i=L, R$. 
Likewise, the probability distribution for the phase difference is calculated as in Refs.~\cite{Barnett1990Dec, Jessop2020Feb} using the phase state basis, $\ket{\theta}=\sum_{n=0}^N e^{i n\theta}\ket{n},$ where $\ket{n}$ are the number states truncated to at most $N$ photons (here we use $N=15$ to ensure converged results):
\begin{equation}
P(\theta)=\!\!\!\sum_{n,m=0}^N\,\sum_{k=\max(n,m)}^N\!\!\!\!\!\frac{e^{i\theta(m-n)}}{2\pi}\bra{n,k-n}\rho_{st}\ket{m,k-m}.
\label{eq:phase_dif_probability}
\end{equation}
This phase difference probability distribution is shown in the last column of Fig.~\ref{fig:two_mode_prob_dist} and seen to display a pronounced peak reflecting phase synchronization. At zero detuning, $\delta\omega=0$, the two oscillators are locked at a phase difference of $\theta=\pi$, consistent with the linear analysis in terms of the Adler equation~\eqref{eq:linear_phase_sync}. Increasing the detuning, the phase locking softens and the phase difference decreases to $\pi/2$. Once again, this behavior is largely consistent with that of the saddle-point equations, and even so for $\delta\omega/\omega_{0}=0.004$, when according to the LME results the limit cycles are already collapsed ($r_W=0$) and yet display some degree of phase locking (cf. last row of Fig.~\ref{fig:two_mode_prob_dist}). We note that for the $L/R$-symmetric parameter set used here, changing the sign of the detuning, $\delta\omega\to-\delta\omega$, corresponds merely to interchanging $L$ and $R$ in Fig.~\ref{fig:two_mode_prob_dist}.  

\begin{figure}[t]
    \centering
    \includegraphics[width=\linewidth]{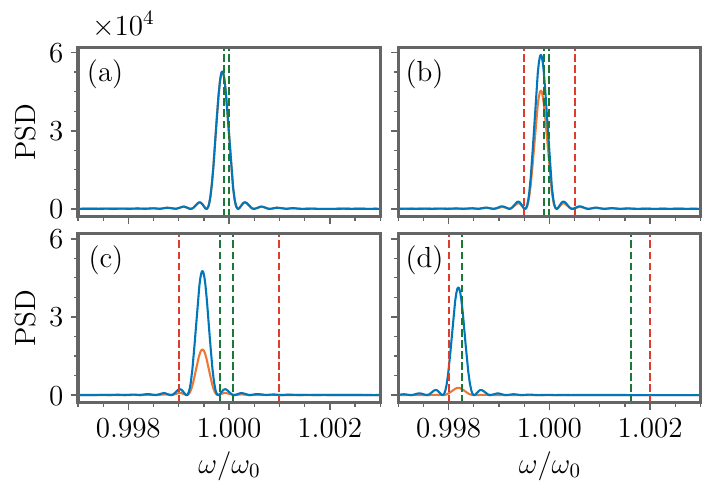}
    \caption{Power spectral density for the left/right mode (orange/blue lines) obtained from the saddle-point solution. The coupling strength is $g_L=g_R=0.045\omega_0.$ The detuning increases from (a) to (d) as $\delta\omega/\omega_0=[0, 0.001, 0.002, 0.004]$. The red and green gridlines show the bare oscillator frequencies, and the real part of the eigenfrequencies of the linear model, respectively}
    \label{fig:PSD_2mode}
\end{figure}

Calculating the PSD from  solutions of the saddle-point equations~\eqref{eq:saddle_point_full}, frequency synchronization is revealed as the merging of PSD maxima at one and the same frequency for the left and right modes observed in Fig.~\ref{fig:PSD_2mode}. As mentioned above, integrating the PSD yields $\langle X^2\rangle,$ and so different peak areas reflect different limit cycle radii for the two modes when $\delta\omega\neq0$, consistent with Fig.~\ref{fig:two_mode_prob_dist}. 
Contrasting this result with the level attraction of the linear model in Fig.~\ref{fig:eigenmodes}, marked by the green dashed lines in Fig.~\ref{fig:two_mode_prob_dist}, one observes an actual frequency synchronization when the nonlinearity is included, as compared to the small but finite separations of normal mode frequencies arising from the level attraction in the linear model. Furthermore, the PSD shows synchronization even for $\delta\omega/\omega_0=0.004$ (cf. Fig.~\ref{fig:two_mode_prob_dist}(d)), which is well outside the region of strong level attraction. This nicely illustrates the fact, that level attraction in the linear model is qualitatively {\it and} quantitatively different from an actual frequency synchronization\cite{Bernier2018Aug, Pikovsky2001Oct}, which is only possible with a nonlinearity.

\section{Conclusions and discussion}

We have analyzed the properties of both a single, and two quantized microwave resonators coupled to a specific tunable electronic environment. The DQD environment sketched in Fig.~\ref{fig:DQD_sketch} is particularly interesting because of its electrical tunability and not least because it has already been shown to function as an effective gain medium at sufficiently large bias voltage when properly gated~\cite{Liu2015Jan}. Summarizing our main findings, we have established that this system serves as a microscopic model of electrons and photons, which under cryogenic conditions ($T\ll\omega_{0}$) realizes quantum limit-cycle oscillators. Depending on their frequency detuning, the two oscillators will undergo quantum synchronization in phase and frequency, as defined respectively by their joint phase-difference probability and individual power-spectral densities. 

Our results were obtained from a perturbative analysis of the nonequilibrium Keldysh action, valid for a small effective fine-structure constant ($g_i\ll\omega_i$) and assuming $\kappa\ll\Gamma,\omega_{i}$ to ensure a time-local action. The semiclassical dynamics was obtained by solving the nonlinear saddle-point equations (SPE), and the effects of quantum fluctuations were studied for a single resonator by means of the corresponding Fokker-Planck equation (FPE). These approximate results were compared with the exact solution of a corresponding Lindblad master equation (LME), taking the system to include both the resonators and the DQD electrons, and the bath to comprise the metallic leads and the ohmic resonator environment with corresponding single-electron, and single-photon jump operators. Strictly speaking, the LME is only justified in the limit of infinite bias voltage, where the electronic bath becomes Markovian. In practice, this limit is already reached as soon as the bias voltage is slightly ($\sim\Gamma$) larger than the electronic excitation energy (as indicated by Figs.~\ref{fig:polarization} and~\ref{fig:lambda}), and is definitely satisfied for $V=10 \omega_{0}$, which we have used throughout. 
Overall, we find excellent agreement between the different calculations (SPE, FPE and LME), and we have indicated where the perturbative method starts deviating markedly from the LME.

Since the aim of this work was to pave the way from a microscopic model of a realistic quantum electronic cQED system to a pair of coupled quantum limit-cycle oscillators, we have deliberately chosen to focus on a single exemplary parameter set, which brings forth the salient features necessary to demonstrate this: self-sustained limit-cycle oscillations, entrainment to external periodic forcing, and frequency and phase synchronization among two slightly detuned resonators. Compared to the works (cf. e.g. Ref.~\onlinecite{Sudler2024May} and references therein), some of which are cited above, which have come to define the emergent research field of quantum synchronization in terms of LME with specific jump operators, a microscopic model like the one considered here imposes severe, yet very physical restrictions on the quantum dynamics. Within a LME approach, the quantum Stuart-Landau oscillator (often not distinguished from the quantum van der Pol oscillator) is usually defined in terms of two or three specific jump operators~\cite{Lee2013Dec, Walter2014Mar}, for which the rates can be tuned individually. Adjusting one of the parameters in our microscopic model, say bias voltage or QD-lead tunneling rates, one might change the one-photon gain while at the same time inadvertently inflicting a one, or two-photon loss or a Hamiltonian Kerr anharmonicity, say. This makes a direct comparison with our results difficult, but as demonstrated by Cottet et al.~\cite{Cottet2020Oct}, the time-local single-mode version of the Keldysh action~\eqref{eq:Seff} may in some cases allow for the  identification of pairs of jump operators and rates defining a corresponding LME for the resonator only. For our system, this mapping would give rise to the following list of jump operators:
\begin{align}
\left\{\left(\hat{a}^{\dagger}, \gamma_{1}\right), \left(\hat{a}^2, \gamma_{2}\right),\left(\hat{a}, \gamma_3\right),\left(\hat{a}^{\dagger2}, \gamma_{4}\right), \left(\hat{a}^{\dagger}\hat{a}, \gamma_{5}\right)\right\},
\end{align}
with corresponding jump rates given by
\begin{align}
\gamma_{1}&=2\kappa n_{B}+\Pi''-\left(\Pi^{K}\right)''/2,\\
\gamma_{2}&=\Lambda_{1}''+\Lambda_{2}''/2-\Lambda_{5}''/4,\nonumber\\
\gamma_3&=2\kappa(1+n_{B})-\Pi''-\left(\Pi^{K}\right)''/2,\nonumber\\
\gamma_{4}&=-\Lambda_{1}''+\Lambda_{2}''/2-\Lambda_{5}''/4,\nonumber\\
\gamma_{5}&=-2\Lambda_{2}'',\nonumber
\end{align}
together with a Kerr (Duffing) anharmonicity contributing to the effective photon Hamiltonian as $\hat{H}_{\rm Kerr}=K\hat{a}^{\dagger2}\hat{a}^{2}$, with strength $K=(\Lambda_{1}'+\Lambda_{3}')/2$. Nevertheless, this mapping between Keldysh action and resonator LME is only valid if $\Lambda_{1}=\Lambda_{3}^{\ast}$ and $\Lambda_{2}'=\Lambda_{4}=0$~\cite{Cottet2020Oct}, which is clearly not the case here (cf. Fig.~\ref{fig:lambda}). Therefore, one should not expect a direct mapping of results from a single-resonator LME to the results obtained above. Expressing the limit-cycle radius found from the saddle-point analysis in terms of these jump rates,
\begin{align}
r_\ast=\sqrt{-\tilde{\kappa}/\Lambda_1''}=\sqrt{(\gamma_1-\gamma_3)/(\gamma_2-\gamma_4)},
\end{align}
one arrives at the physically meaningful result that it 
is given as the ratio between net 1-photon gain and net 2-photon loss, consistent with LME studies of the Stuart-Landau oscillator~\cite{Lee2013Dec, Walter2014Mar}, which usually consider only $\gamma_1$ and $\gamma_2$. Nevertheless, the invalidity of the mapping clearly demonstrates that in spite of the fact that our effective action for a single resonator is time local within our approximations, it is impossible to describe the resonator alone within a Markovian LME. It is only the reduced density matrix of resonator {\it and} DQD electrons which experiences Markovian baths in the infinite-bias limit.

The two resonators in our system are coupled via the electrons on the DQD, and this leads to a linear coupling, $\Pi_{LR}$, which is mainly dissipative. Our synchronization results are largely consistent with the conclusions drawn from Refs.~\cite{Lee2014Feb, Walter2015Jan}, which studied two dissipatively coupled limit-cycle oscillators. Increasing the frequency detuning, synchronization is modified but remains, even after the limit cycles have collapsed. This is seen in Fig.~\ref{fig:two_mode_prob_dist}, where the Wigner functions calculated from the LME and shown in the last row take their maximum at zero radius. 

Typical well-controlled and tunable DQD systems are Coulomb blockaded due to a substantial capacitive charging energy of the individual QDs. This will change the electrical polarizability encoded in $\Pi$ and modify the resonance criterion with the photons, which calls for an entirely different gate setting. Nevertheless, as we demonstrate in Appendix~\ref{app:Coulomb} and as observed experimentally in Ref.~\onlinecite{Liu2014Jul}, the basic mechanism persists and provides for sufficient net energy transfer from electrons to photons to sustain the limit cycle. 

\section{Outlook}

The parameters considered here were chosen to demonstrate the quantum limit-cycle behavior and synchronization. In future work, it would be interesting to explore other parameter regimes. At large voltage bias, the LME can be employed to explore a larger parameter region, using for example the reduced basis method introduced in Ref.~\onlinecite{Christiansen2025Oct}. Staying at weak coupling, the full potential of the Keldysh path integral approach could also be used to explore the non-Markovian regime near $V\sim\varepsilon$, where electronic properties vary the most.

The Coulomb blockade may be partially lifted by increasing the QD-lead tunneling rates, $\Gamma$. This will lead to Kondo correlations, extending the electron polarization clouds further into the metallic leads, which however will degrade with increasing bias voltage. The strongly non-Markovian influence of Kondo correlations on the resonators presents another interesting question to address in future research, possibly starting from the approach taken in Ref.~\onlinecite{Dorda2014Apr}.

Finally, the experimental realization of the system studied here should be within reach of existing superconducting resonators and DQDs in gateable semiconductor nanowires~\cite{Liu2015Jan, Haldar2024Feb, Haldar2025Oct, Janik2025Mar}. A base temperature of $T\approx 10$ mK and a resonator frequency of $\omega_{0}/2\pi\approx 8$ GHz, as in Refs.~\cite{Liu2014Jul, Janik2025Mar}, corresponds to a dimensionless ratio of $T/\omega_{0}\approx 0.03$. This is 30 times larger than the value used throughout this work, deliberately chosen to focus on the effects of quantum fluctuations, indicating that most likely thermal fluctuations cannot be ignored altogether in a real system. Experiments have already demonstrated the limit-cycle behavior~\cite{Liu2015Jan} and entrainment~\cite{Liu2017Nov} of a single resonator coupled to a voltage-biased DQD, and realizing similar systems with two resonators coupled to the same DQD, it should be possible to study the quantum synchronization described here in an actual experiment.

\section{Acknowledgements}
We acknowledge useful discussions with Hans Christiansen and Johan Skak. This work is supported by Novo Nordisk Foundation grant NNF20OC0060019 (CH)

\FloatBarrier

\appendix

\section{PERLind approach to finite-bias system with Coulomb interactions}
\label{app:Coulomb}
\begin{figure}[t]
    \centering
    \includegraphics[width=\linewidth]{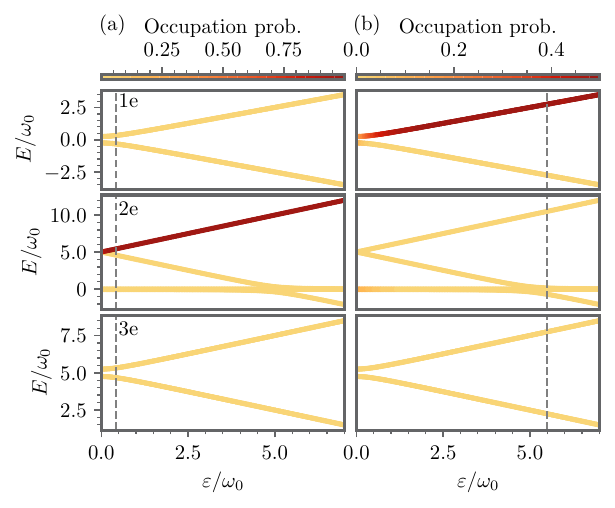}
    \includegraphics[width=\linewidth]{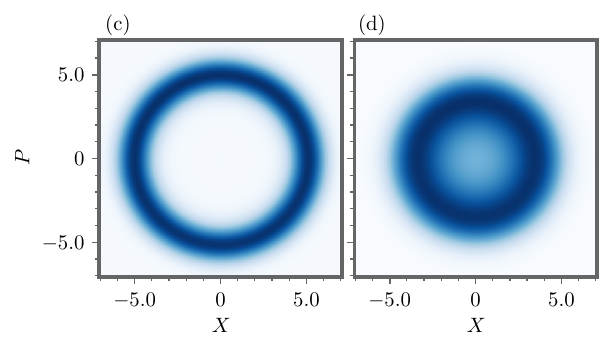}
    \caption{Occupation probabilities of the eigenstates of the interacting DQD system ($g=0$) calculated using PERLind for large, $V=100$ (a) and intermediate, $V=8$ (b) bias. The gridlines show the values of $\varepsilon$ used in (c) and (d), where the Wigner function for the two choices of bias is shown. The other parameters are $t=1/4, U=5, \kappa=0.001, \omega_0=5.5, g=0.3$ and in (c): $\varepsilon=5.48, V=8$, (d) $\varepsilon=0.43, V=100$. In (a) the oscillator frequency is resonant with the excitation energy in the 2e-sector, and in (b) with a transition in the 1e-sector.}
    \label{fig:perlind_wigner}
\end{figure}
At finite bias voltages, the electronic reservoirs are no longer Markovian and one may resort to the phenomenological position and energy resolving Lindblad (PERLind) approach~\cite{Kirsanskas2018Jan}. This connects with the LME results for infinite bias voltage, and provides approximate results at finite bias voltage. As for the LME, it readily allows for spinful electrons and Coulomb interactions on the QDs. In Fig.~\ref{fig:perlind_wigner} we show the resulting many-body spectra for a DQD with respectively 1, 2, or 3 electrons, and with a color code to indicate their nonequilibrium occupations. Both the large (panels a and c), and the small bias case (panels b and d) show a pronounced population inversion, but in different electron number sectors.  The corresponding Wigner functions shown in the lower panels testify to the limit-cycle behavior in the presence of strong Coulomb interactions, even at a moderate bias voltage.

\section{Power spectral density}\label{app:psd}

The power spectral density (PSD) is defined here as the Fourier transform of the autocorrelation function. We employ the Mathematica built-in function "PowerSpectralDensity,"~\cite{Mathematica}, which returns
\begin{equation}
P(\omega)=\sum_{h=-(N-1)}^{N-1}r(h)w\left(\frac{h}{2(N-1)}\right)e^{-ih\omega},
\end{equation}
where the weight function is $w(x)=1$ for $-1/2\leq X\leq 1/2$ and zero otherwise, and the auto-correlation function is obtained as
\begin{equation}
r(h)=\frac{1}{N}\sum_{i=1}^{N-h}\left(X_{i+h}-\langle X\rangle\right)\left(X_{i}-\langle X\rangle\right),
\end{equation}
for a discrete time series $X$. For a process with zero mean (as here), the integral of the PSD is simply the variance $\langle X^2\rangle,$ which relates the limit cycle radius to the PSD as
\begin{equation}
r_\text{sp}^2=2\langle X^2\rangle=\frac{1}{\pi}\int_{-\infty}^\infty \dd \omega P(\omega).
\end{equation}

\bibliography{coupler_prb}

\end{document}